# Polarization and localization of single-photon emitters in hexagonal boron nitride wrinkles

*Donggyu Yim, Mihyang Yu, Gichang Noh, Jieun Lee\*, and Hosung Seo\**

Department of Physics and Department of Energy Systems Research, Ajou University, Suwon, Gyeonggi 16499, Korea

**KEYWORDS**

Hexagonal boron nitride, single-photon emitters, quantum emitters, h-BN wrinkles, solid-state quantum information, quantum nanophotonics

**ABSTRACT**

Color centers in 2-dimensional hexagonal boron nitride (h-BN) have recently emerged as stable and bright single-photon emitters (SPEs) operating at room temperature. In this study, we combine theory and experiment to show that vacancy-based SPEs selectively form at nano-scale wrinkles in h-BN with its optical dipole preferentially aligned to the wrinkle direction. By using density functional theory calculations, we find that the wrinkle's curvature plays a crucial role in localizing vacancy-based SPE candidates and aligning the defect's symmetry plane to the wrinkle direction. By performing optical measurements on SPEs created in h-BN single-crystal flakes, we experimentally confirm the wrinkle-induced generation of SPEs and their polarization alignment to the wrinkle direction. Our results not only provide a new route to controlling the atomic position and the optical property of the SPEs but also revealed the possible crystallographic origin of the SPEs in h-BN, greatly enhancing their potential for use in solid-state quantum photonics and quantum information processing.

**INTRODUCTION**

Solid-state single-photon emitters (SPEs) are essential elements for the realization of integrated quantum photonic technologies such as quantum communication and quantum network[1–3]. Among several promising candidates, defect-derived SPEs from h-BN have recently attracted a great amount of attention owing to its spectral stability, brightness, and room-temperature functionality[4–7]. Since the first discovery in 2015[4], the h-BN SPEs have been extensively investigated to understand the SPEs' optical[8–10], structural[11] and electronic[12–14] properties. Furthermore, the two-dimensional (2D) nature of the host h-BN crystal allows for the SPEs to be integrated in diverse solid-state device structures, enabling to tune the SPEs from h-BN by applying a strain[15–17] or an electric field via Stark effect[18–20], which are useful to generate identical single photons emitted from different sources. It has become also possible to couple the SPEs from h-BN to nano-photonic devices such as waveguides and photonic crystals, which are crucial for developing integrated quantum nano-photonic devices[21–24]. Very recently, optically detected magnetic resonance (ODMR) signals were also reported using several SPEs in h-BN, implying the presence of optically addressable spin quantum bits (qubits) in h-BN, which might be similar to the nitrogen-vacancy spin qubit in diamond[25–28],

The atomic origin of the SPEs, however, have not been completely resolved so far[29–35] as they are normally created in random positions with unknown atomic structure during, e.g. a chemical vapor deposition (CVD) growth or a high-temperature annealing process[11,36]. Several defect models were theoretically suggested, and significant advances have been made toward the identification of the SPEs' atomic origin[26–30,32–35,37,38]. However, no conclusive study exists to confirm their one-to-one relation to the SPEs observed in experiment. Various spectroscopic data were collected and analyzed[10,39] in order to understand the atomic origin of the SPEs. Notably, the orientation of the optical transition dipole of the SPEs has been extensively investigated as a symmetry fingerprint of the SPEs[40]. Most of the SPEs' transition dipole has been found randomly oriented[14,41,42] with no correlation between SPEs. Interestingly, some SPEs were found to show intriguing optical transition behaviors such as presence of intermediate defect states[12,13] and high quantum efficiency from resonant excitation[14]. However, none of the defect models that were reported so far has successfully explained all the observed spectroscopic features of the SPEs.

Deterministic control of the SPE position is also a great challenge to overcome toward the practical application of the h-BN SPEs[43–46]. Several previous studies addressed this issue based on a general tendency that point defects prefer to form at low-symmetry regions such as surfaces, interfaces, or deformed area such as wrinkles, which was also widely observed for SPEs in h-BN[11,18,37,43–45,47–50] and other 2D host materials[51,52]. Proscia *et al.*, fabricated an array of emitters by placing h-BN on top of a patterned substrate with sharp pillars and showed strain-induced activation of single-photon emissions by carrying out optical spectroscopies including Hanbury Brown-Twiss (HBT) correlation measurements[45]. Choi *et al.* created an array of physical holes in h-BN and found that the SPEs are selectively created at the edges of the holes[43]. We note, however, that the microscopic mechanism of the atomic localization of the SPEs in h-BN at such deformed regions is still lacking, which is of great importance toward the fully deterministic control over the position and the optical properties of the h-BN SPEs.

In this study, we show that nano-scale wrinkles in h-BN can be employed to localize SPEs and control their optical property. Furthermore, our combined first-principles theory and optical spectroscopy strongly suggest that the SPEs on the wrinkles are vacancy-derived, significantly enhancing the understanding on the atomic origin of the h-BN SPEs. By using density functional theory, we show that vacancy-based defects gain a significant amount of energy when created on top of a wrinkle due to a larger curvature than in a flat area. We find that the large curvature induces a profound structural relaxation in the defects owing to a dimer reconstruction of their dangling bonds. As an important consequence of the structural relaxation, the symmetry axis of the defects is found to be aligned to the wrinkle direction. Such a characteristic is also evidenced by optical experiments showing an obvious correlation between the polarization orientation of SPEs with the wrinkle direction. Our combined theoretical and experimental study laid down a solid ground for the use of the curvature of h-BN as an effective tool to atomically engineer the SPEs in h-BN at nano-scale.

**EXPERIMENTAL AND THEORETICAL METHODS**

**Density functional theory of defects in curved h-BN.** We performed density functional theory (DFT) calculations using plane-wave basis functions with an energy cutoff of 85 Ry along with optimized norm-conserving Vanderbilt (ONCV) pseudopotentials[53–56] as implemented in the QUANTUM ESPRESSO code[57]. Valence electron configurations used in the pseudopotentials are $2s^22p^1$ for B, $2s^22p^3$ for N, and $2s^22p^2$ for C. We employed the Perdew-Burke-Ernzerhof (PBE) semi-local functional to describe the exchange-correlation potential[58].

To simulate the presence of an isolated point defect in h-BN, we used a supercell method. Each supercell contains 192 atoms and a defect is created in the center of the supercell. The Brillouin zone is sampled with the Γ point only. We tested supercells with different sizes and checked that the supercell size of 192 atoms is enough to separate the central defect and its periodic images.

To study the effect of wrinkle's curvature on defects in h-BN, we employed two different h-BN wrinkle structures with two different curvatures: Gaussian-shaped and sine-shaped. We used the two different wrinkle types in order to cross-check the validity of our theoretical results. Namely, the main conclusion should not change depending on the shape of the wrinkle. The wrinkles are induced by applying compressive uniaxial strain to the supercell: 5% and 10% strains were used, which generate wrinkles with curvatures of 0.75 nm$^{-1}$ and 0.94 nm$^{-1}$, respectively, at the top of the sine-shaped wrinkle, and curvatures of 1.02 nm$^{-1}$ and 1.45 nm$^{-1}$, respectively, for the Gaussian-shaped wrinkle. We find that the response of the h-BN lattice to the compressive strain is dominantly given as curving the plane, maintaining its intrinsic bulk bond length and bond angle the same after structural relaxation of the supercell. Our finding is consistent with a previous theoretical study, which showed that h-BN is flexible under bending and easily curves in the presence of lattice[59]

To shed light on the behavior of SPEs of h-BN in the presence of wrinkles, we considered 6 different point defects: $V_B$, $V_N$, $C_B$, $C_N$, $V_NC_B$, and $V_NN_B$. These defects are intrinsic defects naturally formed during the CVD growth of h-BN and high-temperature annealing of h-BN single crystals[11,21,36]. In particular, $V_NC_B$ and $V_NN_B$ are considered to be strong defect candidates for the SPEs from h-BN due to their deep-level structure consistent with the SPEs' optical emissions[29,32,33]. Additionally, recent study suggested $V_B$ and C-related defects as potential origins of optically detected magnetic resonance (ODMR) signal[25–28].

**Sample preparation: h-BN exfoliation and thermal annealing.** In our experiment, h-BN bulk crystal purchased from HQ Graphene is exfoliated onto SiO$_2$/Si substrate. The exfoliated h-BN is cleaned by acetone and isopropyl alcohol to remove the residue. Then h-BN flakes are annealed at 800 °C for 2 hours in a chamber filled with Ar gas. After annealing, the sample is cooled down to room temperature and the surface morphology measurement is performed using atomic force microscopy (AFM). The h-BN thickness measured from AFM is about 80 nm (Supplementary Information Figure S1).

**Optical studies of the emitters in h-BN.** In photoluminescence (PL) measurement, 532 nm laser is used as an excitation source (P = 200 $\mu$W). The linearly polarized laser is focused onto the sample using an objective lens (NA = 0.60) and the collected light by the same objective is reflected by the beam splitter and guided to enter the spectrometer equipped with a charge-coupled device (CCD). In the collection path, a confocal set-up with a 150 $\mu$m pinhole is used to block the scattered background light and increase the spatial resolution of the 2D scan. All measurements are performed at 10 K.

For the polarization dependent measurement of the excitation laser, we used a half-wave plate (HWP) in the excitation path to rotate the polarization of the laser with respect to the sample orientation. For the emission polarization dependent measurement, the polarization of the laser is fixed, and the emitted light is filtered by an analyzer in the collection path. In order to change the collection polarization while minimizing the influence on the optical path, HWP is included in front of the analyzer and rotated to vary the collection polarization angle.

**RESULTS**

The effect of curved surface of h-BN on the formation of defect centers is first examined experimentally by measuring emitters in exfoliated h-BN flakes with wrinkles. The microscope image of the h-BN flake used in our study is shown in Fig. 1a, which contains several tens of local emitters that are induced by thermal annealing[4]. To identify the location of the emitters, we spatially scanned the 2D emission spectrum of the flake using micro-photoluminescence (µ-PL) set-up at the temperature of 10 K. In the 2D PL map overlapped on Fig. 1a, we present three emitters formed on a h-BN flake, which are indicated by open circles; the emitters A and B are on wrinkle 1 and 2, respectively, and the emitter C is positioned off wrinkle. The height and width of the wrinkle 1 is measured to be around 110 nm and 430 nm, respectively (Supplementary Figure S1). From our measurement, the three emitters are observed to be spatially localized in 2D PL map with narrow zero phonon lines in the emission spectra. Also, dips in the second-order correlation function measurement confirm that these emitters are single photon sources created in h-BN[18].

Then we measured the polarization dependence of the emitted light by varying the excitation laser polarization and collected polarization. Interestingly, for the emitter A on wrinkle 1, the excitation and emission polarizations have the same orientation, which also coincides with the direction of the wrinkle (Fig. 1b). The same relation between the emitter's polarization orientations and the wrinkle direction is also observed for the emitter B on a different wrinkle (wrinkle 2 in Fig. 1c). This observation suggest that the excitation and emission dipole orientations of the curvature-induced emitters preferentially align to the direction of the wrinkle. On the other hand, for the emitter formed on a flat surface (emitter C), the measured excitation and emission polarizations are significantly misaligned from each other (Fig. 1d).

To systematically investigate the correlation between the dipole orientation of the emitters and wrinkles, we analyzed dozens of emitters on the same h-BN flake. Among these, 44 emitters are found on wrinkles and 23 emitters are found on flat surfaces, indicating that more emitters are localized on curved area[45]. Note that we counted only the emitters that produced stable luminescence during the measurement. For the wrinkle-induced emitters, we focus on wrinkle 1 and 2 shown in Fig. 1a, which are showing the angle of 165° and 99° with respect to the x-axis, respectively. By measuring the polarization dependence of these emitters, we found that most of the excitation and emission polarization orientations align to the wrinkle directions within the accuracy of ± 5° (Fig. 2a). Detailed polarization dependence measurement data of one of the emitters is shown in Supplementary Figure S2. However, for the emitters found on flat surface (see Fig. 2b), the dipole orientations of the emitters are randomly distributed and large misalignments between excitation and emission polarizations are frequently found. These experimental observations suggest that the emitters formed on wrinkles have distinctive origins compared to that formed on flat surface.

In addition to the polarization dependence measurement, the wrinkle-driven emitters and emitters on flat surface have contrasting spectral distributions. While the emission energies of the wrinkle-induced emitters are mostly centered around 1.88 or 2.16 eV (Fig. 2c), those of the emitters on flat area are randomly distributed in a wider spectral range (Fig. 2d). More data on random energy distribution of the emitters found on the flat surface is shown in Supplementary Figure S3. For the emitters centered around two different localized energies, previous studies have suggested a few possible intrinsic defect models in h-BN[32,60]. Therefore, exploiting wrinkles in h-BN not only allows controlled fabrication of SPEs in terms of position, optical energies and polarization, it also provides a useful tool to unveil their chemical compositions and crystallographic structures.

To shed light on the apparent correlation between the h-BN SPEs and the wrinkles observed in experiment, we carried out DFT calculations. Figure 3 shows the curvature-driven energy gain of defects as a function of position across a h-BN wrinkle. In Fig. 3, we consider the arm-chair direction of h-BN as the wrinkle direction and we will consider general situations later in this paper. The energy gain is defined as $E_{def}(x) - E_{def}(flat)$, where $E_{def}(x)$ is the total energy of a defect as a function of defect position ($x$) across a h-BN wrinkle and $E_{def}(flat)$ is the energy of the same defect placed at a flat area of h-BN. Figure 3a and 3b show two different wrinkle structures, sine-shaped and Gaussian-shaped, respectively, used in our simulations. We selected two types of intrinsic defects in h-BN, which are vacancy-derived defects ($V_N$, $V_B$, $V_NC_B$, $V_NN_B$) and substitutional impurities ($C_B$ and $C_N$). We remark that vacancy-based defects are considered as important defect models for the h-BN SPEs[4,27,30,32,33,35,38]. The optical energy of $V_NN_B$ was calculated to be around at 590 nm, where many occurrences of SPEs were frequently found[61]. In addition, $V_NC_B$ and $V_B$ (q = -1) were theoretically shown to explain many of the observed spectroscopic features of several h-BN SPEs[25–28,38].

Fig. 3c and 3d show the curvature-driven energy gains of the defects placed at sine-shaped wrinkles with curvature of 0.75 nm$^{-1}$ and 0.94 nm$^{-1}$, respectively, at their wrinkle top. In Fig. 3c, we find that the energy gains for all the vacancy-derived defects increase as the defects approach the wrinkle and the largest energy gain is achieved at the top of the wrinkle where the curvature is maximized: 0.49 eV, 0.56 eV, 0.67 eV, and 0.68 eV for $V_N$, $V_NC_B$, $V_NN_B$, and $V_B$, respectively. For the substitutional C impurities, however, a negligible energy gain is obtained across the wrinkle. Furthermore, as shown in Fig. 3d, the energy gain for the vacancy-based defects is further increased when the curvature at the wrinkle top is increased by 25%: 30%~40% increase for $V_N$, $V_NC_B$, and $V_NN_B$ and 70% increase for $V_B$ (see Supplementary Figure S4). Our results show that the wrinkle's curvature is the main cause of the energy gain. On the other hand, the negligible wrinkle-driven energy gain for the substitutional C defects remains the same although the wrinkle's curvature is increased.

To cross-check our conclusion obtained with the sine-shaped wrinkles, we performed the same calculations, but using Gaussian-shaped wrinkles having two different curvatures, 1.02 nm$^{-1}$ and 1.45 nm$^{-1}$ at the wrinkle top, and the results are summarized in Fig. 3e and 3f, respectively. We find that the vacancy-derived defects gain the maximum energy when formed on top of the wrinkle, which ranges from 0.71 eV to 1.26 eV and from 0.91 eV to 1.75 eV in Fig. 3e and 3f, respectively. In addition, by comparing the energy gains in Fig. 3f to those in Fig. 3e, we note that 42% increase of the wrinkle's curvature leads to 25% ~ 40% increase in the energy gain for $V_N$, $V_NC_B$, $V_NN_B$, and $V_B$. The substitutional C impurities, however, gain negligible energy in the proximity of the Gaussian-shaped wrinkles regardless of the curvature at the wrinkle top. Our results obtained with both sine-shaped and Gaussian-shaped wrinkles with varying curvatures (see Supplementary Figure S4) clearly demonstrate that the formation of vacancy-derived defects is much easier in wrinkles than in a flat h-BN plane owing to the curvature-induced large energy gain lowering the defect formation energy.

In Fig. 4, we investigate the vacancy-derived defects created on top of h-BN wrinkles that are randomly misaligned with respect to the crystallographic armchair direction of the h-BN lattice. We find that the curvature-driven energy gain of the vacancy-based defects is maximized when their $C_S$ mirror plane is parallel to the wrinkle direction. Fig. 4a shows three possible angles ($\theta_{1,2,3}$) for a $V_NX_B$-type defect (X = C, N) with respect to an arbitrary oriented wrinkle direction. We remark that the largest and smallest defect angles possible are 90° and 0°, respectively (see Supplementary Figure S5). In Fig. 4b, we compare the curvature-driven energy gain of the $V_NN_B$ and $V_NC_B$ defects as a function of the defect angle with respect to the wrinkle direction. We observe that for both defects the energy gain monotonically increases by ~ 1 eV as the defect angle is decreased from the maximum angle to the minimum angle. Our results show that the curvature of a h-BN wrinkle not only drives atomic localization of the vacancy-derived defects but determines the orientation of the defect geometry with respect to the wrinkle.

We now turn to the analysis of change in the electronic structure of the defects occurring in wrinkles. We find that the microscopic origin of the wrinkle-driven localization and orientation of the vacancy-derived defects is traced to dimerization of atoms in the defects, whose strength is maximized at the top of wrinkles. For the $V_NN_B$ and $V_NC_B$ defects in a flat h-BN plane, the two nearest neighboring B atoms of the $V_N$ site form a dimer due to the interaction between their dangling bonds (see Supplementary Figure S6). The dimer length is calculated to be 1.97 Å and 1.97 Å for $V_NN_B$ and $V_NC_B$, respectively. When the $V_NN_B$ and $V_NC_B$ defects are created at the wrinkle top, however, their dimer length is significantly shortened by ~0.2 Å compared to their dimer length in a flat area as shown in Fig. 4 (c). Furthermore, we find that the dimer length is monotonically decreased as the defect angle is decreased, which explains the increase in the energy gain shown in Fig. 4c. Our result shows that the curved environment of a wrinkle is favorable to vacancy-based defects as it enables an enhanced dimerization, thus significantly lowering its energy.

As a consequence of the defect alignment at a wrinkle top discussed above, we find that the optical transition dipole of the $V_NX_B$ (X=C, N) defects is also aligned to the wrinkle direction. The point group symmetry of the $V_NX_B$ defects in the middle of a wrinkle is $C_S$, whose mirror plane is perpendicular to the dimer direction of the defects and the h-BN plane. According to the group theory[40], the possible optical transitions for the $V_NN_B$ and $V_NC_B$ defects are A' – A', A' – A'', and A'' – A''. The lowest energy transition is known to be A' to A'[17], which has X and Y dipoles. The X dipole is perpendicular to the dimer direction, thus aligned to the wrinkle direction. To verify the group theory result, we also numerically calculate the transition dipole of the $V_NN_B$ and $V_NC_B$ defects between the A' ground state $(a'(1)^2a'(2)^1)$ and the A' excited state $(a'(1)^1a'(2)^2)$ using the Kohn-Sham defect orbitals[62]. As shown in Fig. 4d, the dipole is aligned to the wrinkle direction for light emission or absorption occurring perpendicular to the h-BN plane, which is consistent with the group theory result and our experimental observation.

**CONCLUSION**

In conclusion, we used a combined theoretical and experimental method to investigate the intimate coupling of wrinkles and single photon emitters in h-BN. We experimentally observed that the number of emitters found on wrinkles far exceed the number of emitters found on flat plane, considering the density of emitters found from the same unit area. In addition, these emitters on wrinkles show a characteristic energy distribution, which is centered around 1.88 eV and 2.16 eV, implying that the emitters formed on wrinkles have a unique origin compared to those found on flat region. Using density functional theory, we demonstrated that the vacancy-derived defects can lower their defect formation energy significantly if created in wrinkles owing to the curvature-driven energy gain. On the other hands, substitutional impurities exhibited no energy gain in the proximity of wrinkles.

Furthermore, we experimentally observed that the emitters found on wrinkled area show clear dipole orientations for both excitation and emission, which align to the wrinkle directions. Such dependence indicates clear anisotropy in the emitter's crystal structure. Theoretically, we discovered that the most stable atomic arrangement of the vacancy-derived defects considered in this study is one with the defects' $C_S$ mirror plane being parallel to the wrinkle direction. Due to the aligned defect's symmetry plane, the defect's optical dipole is also oriented to the wrinkle direction, which is in an excellent agreement with the experimental observation. Based on our numerical results and group theory analysis, the atomic origin of the SPEs formed in wrinkles is considerably narrowed down to vacancy-based axial defects such as $V_NN_B$ and $V_NC_B$.

We should mention, however, that it is still not possible to specify the exact atomic structure of the SPEs in h-BN wrinkles solely based on our study. Furthermore, as shown in Fig. 2(c), the emission energy of the SPEs in h-BN wrinkles shows distributions about two central peaks, suggesting the possible existence of various defect types. Based on our study, other vacancy-including axial defects could also easily form in h-BN wrinkles owing to the curvature-driven energy gain. Such examples may include $V_NO_{2B}$, $V_NO_B$, $V_BC_N$, $V_NC_{2B}$ to name a few. Furthermore, previous studies demonstrated that the emission energy of h-BN SPEs is highly sensitive to local strain environment or its charge state[15-18,20,63], which may contribute to the broadened spectral range of SPEs in h-BN wrinkles. These issues are, however, beyond the scope of this study and we leave them for future study.

Our combined experimental and theoretical work lays a solid foundation to understand and utilize the profound effects imposed by wrinkles on the h-BN SPEs. We found that the dimerization of atoms in vacancy-derived defects plays a key role in constraining the atomic and optical property of the defects in wrinkles. As dangling bonds and their interaction leading to dimerization of atoms are inherent to any vacancy-derived defects in h-BN, the essential physics revealed in this study should apply to other possible vacancy-based SPEs with complex structures[30,33,34]. In addition, it is worth mentioning that controlled folding methods are being actively developed for various 2D materials systems[64] including h-BN[65]. Combining controlled folding of h-BN with curvature-driven creation and control of vacancy-based SPEs would be a promising route to deterministic generation and manipulation of single-photon emitters in h-BN.

**FIGURES**

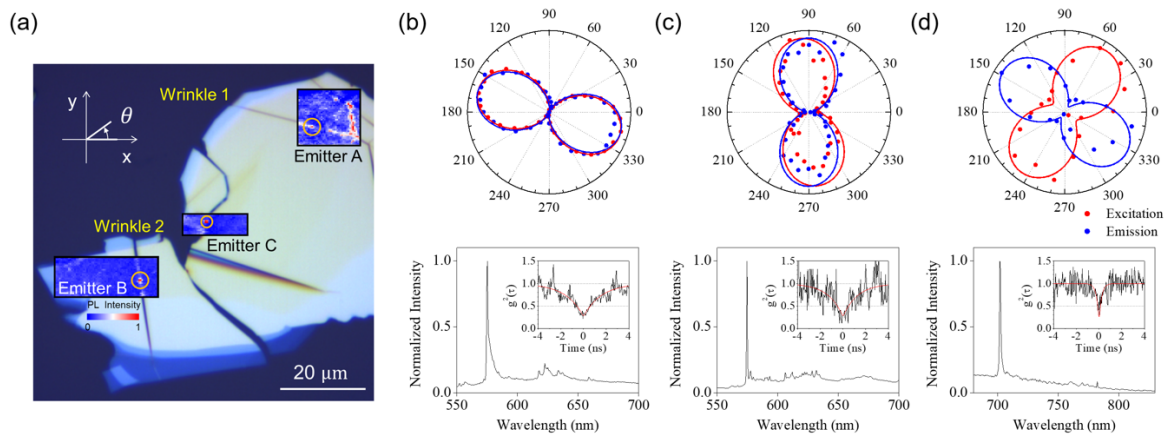

**Figure 1**. Optical measurements and polarization dependence of h-BN emitters. (a) Microscope image of the h-BN flake used in our experiment. The 2D PL map of the emitter A, B, and C are overlapped on the microscope image. (b-d) Polar plot of the excitation and emission polarization dependence (upper) and emission spectrum (lower) of emitter A on wrinkle 1 (b), emitter B on wrinkle 2 (c), and emitter C on flat surface (d). The angle θ of the polarization is measured from the x-axis shown in (a). The polar plots are fitted by $I(\theta) = A\cos^2(\theta - \theta_0) + B$ to find the polarization angle, $\theta_0$. In the inset of the lower panels, the second-order correlation measurement data (black solid line) is shown and fitted by $g^2(\tau) = 1 - e^{-|\tau|/\tau_1}$ convoluted with the detector's response function where the emitter's lifetime, $\tau_1$, is the fitting parameter; $\tau_{1,A} = 1.55$ ns $\tau_{1,B} = 1.28$ ns and $\tau_{1,C} = 0.22$ ns (red solid line).

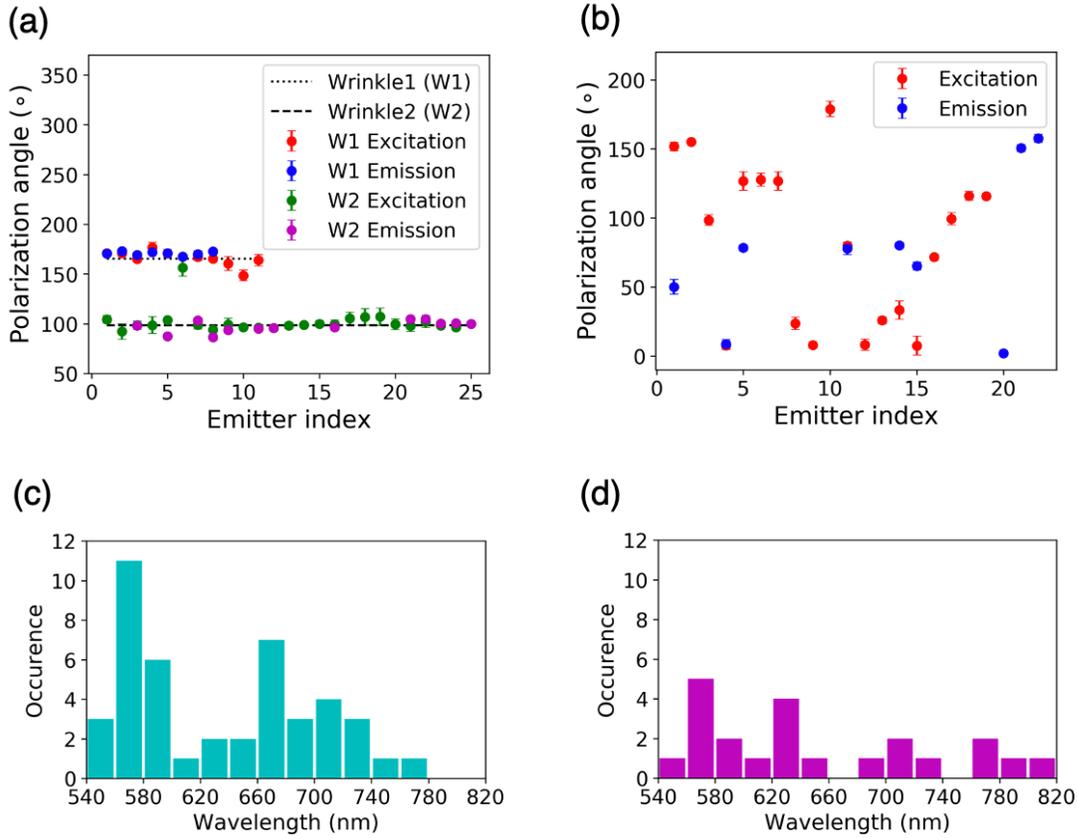

**Figure 2.** Statistical studies on h-BN emitters. (a) The excitation and emission polarization angle $\theta_0$ of the emitters formed on wrinkle 1 (W1) and wrinkle 2 (W2). The angle of W1 and W2 are indicated by the short and long dashed lines, respectively. (b) The excitation and emission polarization angle $\theta_0$ of the emitters formed on flat surface. (c) Histogram of the center wavelength of the emitters formed on wrinkles. (d) Histogram of the center wavelength of the emitters formed on flat surface.

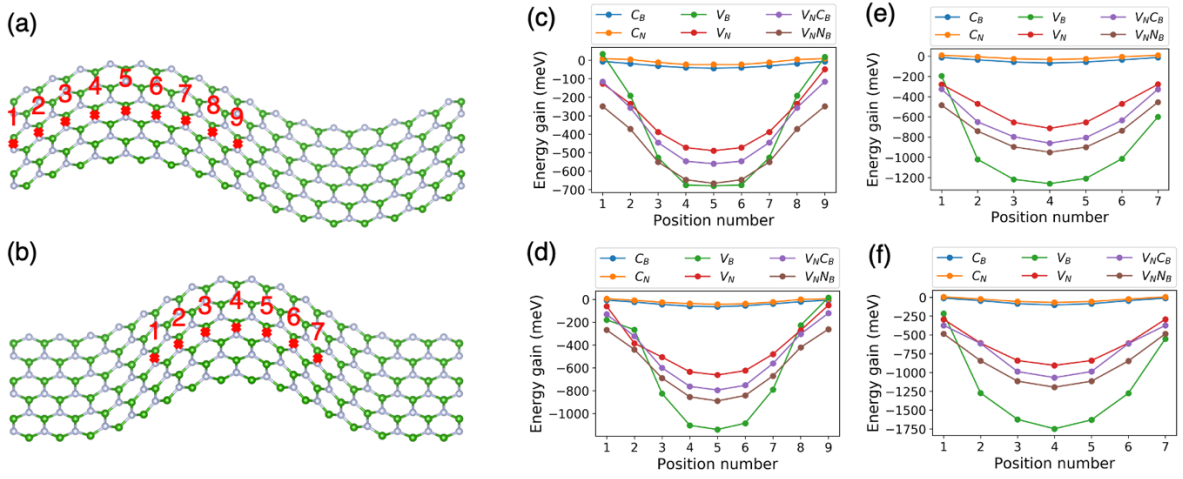

**Figure 3.** Wrinkle-driven atomic localization of single-photon emitters in h-BN. (a, b) Sine-shaped (a) and Gaussian-shaped (b) h-BN supercells used for simulations of defects in the presence of wrinkle. 9 and 7 lattice sites are considered as a possible defective site across the wrinkle for the sine-shaped and Gaussian-shaped supercells, respectively. The considered sites are denoted with red asterisks. (c, d) Curvature-driven energy gain of various defects (see the text) as a function of the position across the sine-shaped wrinkle with two different curvatures: 0.75 nm$^{-1}$ and 1.05 nm$^{-1}$ in (c) and (d), respectively. (e, f) Curvature-driven energy gain of the same defect models as a function of the position across the Gaussian-shaped h-BN wrinkle with two different curvatures: 1.08 nm$^{-1}$ and 1.58 nm$^{-1}$ in (e) and (f), respectively.

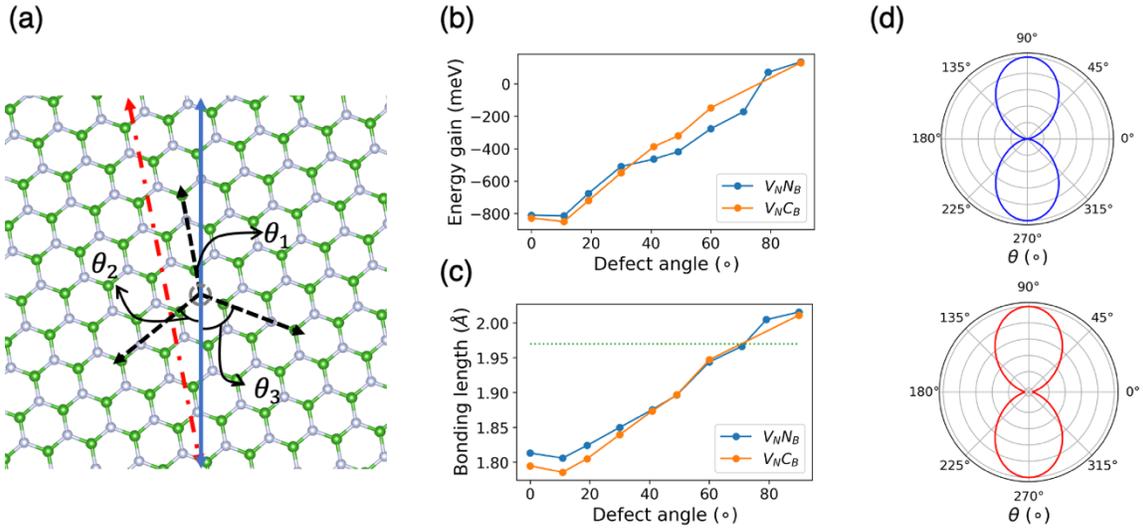

**Figure 4.** Alignment of the $V_NX_B$-type defects with the wrinkle direction. (a) Three possible defect angles ($\theta_1, \theta_2, \theta_3$), which a $V_NX_B$ defect can have with respect to a wrinkle. A possible direction of a wrinkle is indicated by a solid arrow and the crystallographic arm-chair direction is denoted by a red dotdashed arrow. For a $V_NX_B$-type axial defect, three crystallographic directions are possible, which are denoted with black dashed arrows. The smallest possible defect angle with respect to the wrinkle direction is $\theta_1$, which the same as the angle between the armchair direction and the wrinkle direction. (b) The energy gain of the $V_NN_B$ and $V_NC_B$ defects (see the main text for the definition) as a function of the defect angle. The zero energy is the energy of the defects in a flat h-BN area. (c) The dimer length of the $V_NN_B$ and $V_NC_B$ defects as a function of the defect angle. The dotted line indicates the dimer length of the defects in a flat h-BN plane. (d) The calculated polarization dependence of $V_NC_B$ (top) and $V_NN_B$ (bottom) created at the top of a wrinkle for light emission or absorption occurring perpendicular to the h-BN plane.

## ASSOCIATED CONTENT

**Supporting Information**. Supplementary Figure S1. Atomic force microscopy (AFM) of wrinkled h-BN. Supplementary Figure S2. Detailed measurement results of additional wrinkle-associated h-BN emitter. Supplementary Figure S3. The histogram of all observed emitters counted from 2D photoluminescence scan map. Supplementary Figure S4. The energy gain of vacancy-derived defects as function of curvature. Supplementary Figure S5. Possible orientations of axial point defects in h-BN wrinkles. Supplementary Figure S6. Electronic structure and defect states of $V_N N_B$ defects.

## AUTHOR INFORMATION

**Corresponding Authors**
*E-mail: jelee@ajou.ac.kr
*E-mail: hseo2017@ajou.ac.kr

**Author Contributions**
D.Y. performed the theoretical calculations. M.Y. and G.N. carried out the optical experiments. J.L. and H.S. supervised the project. All authors contributed to the data analysis and production of the manuscript.

**Notes**
The authors declare no competing financial interest.

## ACKNOWLEDGMENT

This work was supported by the National Research Foundation of Korea (NRF) grant funded by the Korea government (MSIT) (No. 2018R1C1B6008980, No. 2018R1A4A1024157, and No. 2019M3E4A1078666). M.Y., G.N., and J.L. were supported by the National Research

Foundation of Korea (Grants No. 2017R1C1B2002631, No. 2020R1A5A6052558, and No. 2020R1A2C2011334).

**ABBREVIATIONS**

SPE, single photon emitter; h-BN, hexagonal boron nitride; DFT, density functional theory; PL, photoluminescence.

**REFERENCES**

(1) O'Brien, J. L.; Furusawa, A.; Vučković, J. Photonic Quantum Technologies. *Nat. Photonics* **2009**, *3* (12), 687–695. https://doi.org/10.1038/nphoton.2009.229.

(2) Atatüre, M.; Englund, D.; Vamivakas, N.; Lee, S.-Y.; Wrachtrup, J. Material Platforms for Spin-Based Photonic Quantum Technologies. *Nat. Rev. Mater.* **2018**, *3* (5), 38–51. https://doi.org/10.1038/s41578-018-0008-9.

(3) Wang, J.; Sciarrino, F.; Laing, A.; Thompson, M. G. Integrated Photonic Quantum Technologies. *Nat. Photonics* **2019**, *14*, 273–284. https://doi.org/10.1038/s41566-019-0532-1.

(4) Tran, T. T.; Bray, K.; Ford, M. J.; Toth, M.; Aharonovich, I. Quantum Emission from Hexagonal Boron Nitride Monolayers. *Nat. Nanotechnol.* **2015**, *11* (1), 37–41. https://doi.org/10.1038/nnano.2015.242.

(5) Tran, T. T.; Zachreson, C.; Berhane, A. M.; Bray, K.; Sandstrom, R. G.; Li, L. H.; Taniguchi, T.; Watanabe, K.; Aharonovich, I.; Toth, M. Quantum Emission from Defects in Single-Crystalline Hexagonal Boron Nitride. *Phys. Rev. Appl.* **2016**, *5* (3), 034005. https://doi.org/10.1103/physrevapplied.5.034005.

(6) Martínez, L. J.; Pelini, T.; Waselowski, V.; Maze, J. R.; Gil, B.; Cassabois, G.; Jacques, V. Efficient Single Photon Emission from a High-Purity Hexagonal Boron Nitride Crystal. *Phys. Rev. B* **2016**, *94* (12), 121405. https://doi.org/10.1103/physrevb.94.121405.

(7) Sontheimer, B.; Braun, M.; Nikolay, N.; Sadzak, N.; Aharonovich, I.; Benson, O. Photodynamics of Quantum Emitters in Hexagonal Boron Nitride Revealed by Low Temperature Spectroscopy. *Phys. Rev. B* **2017**, *96* (12), 121202. https://doi.org/10.1103/physrevb.96.121202.

(8) Bourrellier, R.; Meuret, S.; Tararan, A.; Stéphan, O.; Kociak, M.; Tizei, L. H. G.; Zobelli, A. Bright UV Single Photon Emission at Point Defects in *h*-BN. *Nano Lett.* **2016**, *16* (7), 4317–4321. https://doi.org/10.1021/acs.nanolett.6b01368.

(9) Exarhos, A. L.; Hopper, D. A.; Grote, R. R.; Alkauskas, A.; Bassett, L. C. Optical Signatures of Quantum Emitters in Suspended Hexagonal Boron Nitride. *ACS Nano* **2016**, *11* (3), 3328–3336. https://doi.org/10.1021/acsnano.7b00665.

(10) Tran, T. T.; Elbadawi, C.; Totonjian, D.; Lobo, C. J.; Grosso, G.; Moon, H.; Englund, D. R.; Ford, M. J.; Aharonovich, I.; Toth, M. Robust Multicolor Single Photon Emission from Point Defects in Hexagonal Boron Nitride. *ACS Nano* **2016**, *10* (8), 7331–7338. https://doi.org/10.1021/acsnano.6b03602.

(11) Chejanovsky, N.; Rezai, M.; Paolucci, F.; Kim, Y.; Rendler, T.; Rouabeh, W.; Oliveira, F. F. de; Herlinger, P.; Denisenko, A.; Yang, S.; Gerhardt, I.; Finkler, A.; Smet, J. H.; Wrachtrup, J. Structural Attributes and Photo-Dynamics of Visible Spectrum Quantum Emitters in Hexagonal Boron Nitride. *Nano Lett.* **2016**, *16* (11), 7037–7045. https://doi.org/10.1021/acs.nanolett.6b03268.

(12) Shotan, Z.; Jayakumar, H.; Considine, C. R.; Mackoit, M.; Fedder, H.; Wrachtrup, J.; Alkauskas, A.; Doherty, M. W.; Menon, V. M.; Meriles, C. A. Photoinduced Modification of Single-Photon Emitters in Hexagonal Boron Nitride. *ACS Photonics* **2016**, *3* (12), 2490–2496. https://doi.org/10.1021/acsphotonics.6b00736.

(13) Schell, A. W.; Tran, T. T.; Takashima, H.; Takeuchi, S.; Aharonovich, I. Non-Linear Excitation of Quantum Emitters in Hexagonal Boron Nitride Multiplayers. *APL Photonics* **2016**, *1* (9), 091302. https://doi.org/10.1063/1.4961684.

(14) Kianinia, M.; Bradac, C.; Sontheimer, B.; Wang, F.; Tran, T. T.; Nguyen, M.; Kim, S.; Xu, Z.-Q.; Jin, D.; Schell, A. W.; Lobo, C. J.; Aharonovich, I.; Toth, M. All-Optical Control and Super-Resolution Imaging of Quantum Emitters in Layered Materials. *Nat. Commun.* **2018**, *9* (1), 874. https://doi.org/10.1038/s41467-018-03290-0.

(15) Grosso, G.; Moon, H.; Lienhard, B.; Ali, S.; Efetov, D. K.; Furchi, M. M.; Jarillo-Herrero, P.; Ford, M. J.; Aharonovich, I.; Englund, D. Tunable and High Purity Room-Temperature Single Photon Emission from Atomic Defects in Hexagonal Boron Nitride. *Nat. Commun.* **2016**, *8* (1), 705. https://doi.org/10.1038/s41467-017-00810-2.

(16) Mendelson, N.; Doherty, M.; Toth, M.; Aharonovich, I.; Tran, T. T. Strain Engineering of Quantum Emitters in Hexagonal Boron Nitride. *arXiv:1911.08072* **2019**.

(17) Li, S.; Chou, J.-P.; Hu, A.; Plenio, M. B.; Udvarhelyi, P.; Thiering, G.; Abdi, M.; Gali, A. Giant Shift upon Strain on the Fluorescence Spectrum of $V_NN_B$ Color Centers in *h*-BN. *arXiv:2001.02749* **2020**.

(18) Noh, G.; Choi, D.; Kim, J.-H.; Im, D.-G.; Kim, Y.-H.; Seo, H.; Lee, J. Stark Tuning of Single-Photon Emitters in Hexagonal Boron Nitride. *Nano Lett.* **2018**, *18* (8), 4710–4715. https://doi.org/10.1021/acs.nanolett.8b01030.

(19) Nikolay, N.; Mendelson, N.; Sadzak, N.; Böhm, F.; Tran, T. T.; Sontheimer, B.; Aharonovich, I.; Benson, O. Very Large and Reversible Stark Shift Tuning of Single Emitters in Layered Hexagonal Boron Nitride. *Phys. Rev. Applied* **2018**, *11* (4), 041001. https://doi.org/10.1103/physrevapplied.11.041001.

(20) Xia, Y.; Li, Q.; Kim, J.; Bao, W.; Gong, C.; Yang, S.; Wang, Y.; Zhang, X. Room-Temperature Giant Stark Effect of Single Photon Emitter in van der Waals Material. *Nano Lett.* **2019**, *19* (10), 7100–7105. https://doi.org/10.1021/acs.nanolett.9b02640.

(21) Tran, T. T.; Wang, D.; Xu, Z.-Q.; Yang, A.; Toth, M.; Odom, T. W.; Aharonovich, I. Deterministic Coupling of Quantum Emitters in 2D Materials to Plasmonic Nanocavity Arrays. *Nano Lett.* **2017**, *17* (4), 2634–2639. https://doi.org/10.1021/acs.nanolett.7b00444.

(22) Kim, S.; Fröch, J. E.; Christian, J.; Straw, M.; Bishop, J.; Totonjian, D.; Watanabe, K.; Taniguchi, T.; Toth, M.; Aharonovich, I. Photonic Crystal Cavities from Hexagonal Boron Nitride. *Nat. Commun.* **2018**, *9* (1), 2623. https://doi.org/10.1038/s41467-018-05117-4.

(23) Caldwell, J. D.; Aharonovich, I.; Cassabois, G.; Edgar, J. H.; Gil, B.; Basov, D. N. Photonics with Hexagonal Boron Nitride. *Nat. Rev. Mater.* **2019**, *4* (8), 552–567. https://doi.org/10.1038/s41578-019-0124-1.

(24) Fröch, J. E.; Kim, S.; Mendelson, N.; Kianinia, M.; Toth, M.; Aharonovich, I. Coupling Hexagonal Boron Nitride Quantum Emitters to Photonic Crystal Cavities. *ACS Nano* **2020**, *14* (6), 7085–7091. https://doi.org/10.1021/acsnano.0c01818.

(25) Chejanovsky, N.; Mukherjee, A.; Kim, Y.; Denisenko, A.; Finkler, A.; Taniguchi, T.; Watanabe, K.; Dasari, D. B. R.; Smet, J. H.; Wrachtrup, J. Single Spin Resonance in a van der Waals Embedded Paramagnetic Defect. *arXiv:1906.05903* **2019**.

(26) Gottscholl, A.; Kianinia, M.; Soltamov, V.; Orlinskii, S.; Mamin, G.; Bradac, C.; Kasper, C.; Krambrock, K.; Sperlich, A.; Toth, M.; Aharonovich, I.; Dyakonov, V. Initialization and Read-out of Intrinsic Spin Defects in a van der Waals Crystal at Room Temperature. *Nat. Mater.* **2020**, 19, 540. https://doi.org/10.1038/s41563-020-0619-6.

(27) Ivády, V.; Barcza, G.; Thiering, G.; Li, S.; Hamdi, H.; Legeza, Ö.; Chou, J.-P.; Gali, A. Ab Initio Theory of the Negatively Charged Boron Vacancy Qubit in Hexagonal Boron Nitride. *npj Comput. Mater.* **2019**, *6* (1), 41. https://doi.org/10.1038/s41524-020-0305-x.

(28) Mendelson, N.; Chugh, D.; Cheng, T. S.; Gottscholl, A.; Long, H.; Mellor, C. J.; Zettl, A.; Dyakonov, V.; Beton, P. H.; Novikov, S. V.; Jagadish, C.; Tan, H. H.; Toth, M.; Bradac, C.; Aharonovich, I. Identifying Carbon as the Source of Visible Single Photon Emission from Hexagonal Boron Nitride. *arXiv:2003.00949* **2020**.

(29) Huang, B.; Lee, H. Defect and Impurity Properties of Hexagonal Boron Nitride: A First-Principles Calculation. *Phys. Rev. B* **2012**, *86* (24), 245406. https://doi.org/10.1103/physrevb.86.245406.

(30) Tawfik, S. A.; Ali, S.; Fronzi, M.; Kianinia, M.; Tran, T. T.; Stampfl, C.; Aharonovich, I.; Toth, M.; Ford, M. J. First-Principles Investigation of Quantum Emission from hBN Defects. *Nanoscale* **2017**, *9* (36), 13575–13582. https://doi.org/10.1039/c7nr04270a.

(31) Wu, F.; Andrew, G.; Sundararaman, R.; Rocca, D.; Ping, Y. First-Principles Engineering of Charged Defects for Two-Dimensional Quantum Technologies. *Phys. Rev. Mater.* **2017**, *1* (7), 071001. https://doi.org/10.1103/physrevmaterials.1.071001.

(32) Abdi, M.; Chou, J.-P.; Gali, A.; Plenio, M. B. Color Centers in Hexagonal Boron Nitride Monolayers: A Group Theory and Ab Initio Analysis. *ACS Photonics* **2018**, *5* (5), 1967–1976. https://doi.org/10.1021/acsphotonics.7b01442.

(33) Sajid, A.; Reimers, J. R.; Ford, M. J. Defect States in Hexagonal Boron Nitride: Assignments of Observed Properties and Prediction of Properties Relevant to Quantum Computation. *Phys. Rev. B* **2018**, *97* (6), 064101. https://doi.org/10.1103/physrevb.97.064101.

(34) Weston, L.; Wickramaratne, D.; Mackoit, M.; Alkauskas, A.; Van de Walle, C. G. Native Point Defects and Impurities in Hexagonal Boron Nitride. *Phys. Rev. B* **2018**, *97* (21), 214104. https://doi.org/10.1103/physrevb.97.214104.

(35) Turiansky, M. E.; Alkauskas, A.; Bassett, L. C.; Van de Walle, C. G. Dangling Bonds in Hexagonal Boron Nitride as Single-Photon Emitters. *Phys. Rev. Lett.* **2019**, *123* (12), 127401. https://doi.org/10.1103/physrevlett.123.127401.

(36) Mendelson, N.; Xu, Z.-Q.; Tran, T. T.; Kianinia, M.; Scott, J.; Bradac, C.; Aharonovich, I.; Toth, M. Engineering and Tuning of Quantum Emitters in Few-Layer Hexagonal Boron Nitride. *ACS Nano* **2019**, *13* (3), 3132–3140. https://doi.org/10.1021/acsnano.8b08511.

(37) Li, X.; Shepard, G. D.; Cupo, A.; Camporeale, N.; Shayan, K.; Luo, Y.; Meunier, V.; Strauf, S. Nonmagnetic Quantum Emitters in Boron Nitride with Ultranarrow and Sideband-Free Emission Spectra. *ACS Nano* **2017**, *11* (7), 6652–6660. https://doi.org/10.1021/acsnano.7b00638.

(38) Sajid, A.; Thygesen, and K. S. $V_NC_B$ Defect as Source of Single Photon Emission from Hexagonal Boron Nitride. *arXiv:2003.00949* **2020**.

(39) Jungwirth, N. R.; Calderon, B.; Ji, Y.; Spencer, M. G.; Flatté, M. E.; Fuchs, G. D. Temperature Dependence of Wavelength Selectable Zero-Phonon Emission from Single Defects in Hexagonal Boron Nitride. *ACS Nano* **2016**, *16* (10), 6052–6057. https://doi.org/10.1021/acs.nanolett.6b01987.

(40) Jungwirth, N. R.; Chang, H.-S.; Jiang, M.; Fuchs, G. D. Polarization Spectroscopy of Defect-Based Single Photon Sources in ZnO. *ACS Nano* **2015**, *10* (1), 1210–1215. https://doi.org/10.1021/acsnano.5b06515.

(41) Jungwirth, N. R.; Fuchs, G. D. Optical Absorption and Emission Mechanisms of Single Defects in Hexagonal Boron Nitride. *Phys. Rev. Lett.* **2017**, *119* (5), 057401. https://doi.org/10.1103/physrevlett.119.057401.

(42) Exarhos, A. L.; Hopper, D. A.; Patel, R. N.; Doherty, M. W.; Bassett, L. C. Magnetic-Field-Dependent Quantum Emission in Hexagonal Boron Nitride at Room Temperature. *Nat. Commun.* **2018**, *10* (1), 222. https://doi.org/10.1038/s41467-018-08185-8.

(43) Choi, S.; Tran, T. T.; Elbadawi, C.; Lobo, C.; Wang, X.; Juodkazis, S.; Seniutinas, G.; Toth, M.; Aharonovich, I. Engineering and Localization of Quantum Emitters in Large Hexagonal Boron Nitride Layers. *ACS Appl. Mater. Interfaces* **2016**, *8* (43), 29642–29648. https://doi.org/10.1021/acsami.6b09875.

(44) Chejanovsky, N.; Kim, Y.; Zappe, A.; Stuhlhofer, B.; Taniguchi, T.; Watanabe, K.; Dasari, D. B. R.; Finkler, A.; Smet, J. H.; Wrachtrup, J. Quantum Light in Curved Low Dimensional Hexagonal Boron Nitride Systems. *Sci. Rep.* **2017**, *7* (1), 14758. https://doi.org/10.1038/s41598-017-15398-2.

(45) Proscia, N. V.; Shotan, Z.; Jayakumar, H.; Reddy, P.; Cohen, C.; Dollar, M.; Alkauskas, A.; Doherty, M.; Meriles, C. A.; Menon, V. M. Near-Deterministic Activation of Room-Temperature Quantum Emitters in Hexagonal Boron Nitride. *Optica* **2018**, *5* (9), 1128. https://doi.org/10.1364/optica.5.001128.

(46) Ziegler, J.; Klaiss, R.; Blaikie, A.; Miller, D.; Horowitz, V. R.; Alemán, B. J. Deterministic Quantum Emitter Formation in Hexagonal Boron Nitride via Controlled Edge Creation. *Nano Lett.* **2019**, *19* (3), 2121–2127. https://doi.org/10.1021/acs.nanolett.9b00357.

(47) Xu, Z.-Q.; Elbadawi, C.; Tran, T. T.; Kianinia, M.; Li, X.; Liu, D.; Hoffman, T. B.; Nguyen, M.; Kim, S.; Edgar, J. H.; Wu, X.; Song, L.; Ali, S.; Ford, M.; Toth, M.; Aharonovich, I. Single Photon Emission from Plasma Treated 2D Hexagonal Boron Nitride. *Nanoscale* **2018**, *10* (17), 7957–7965. https://doi.org/10.1039/c7nr08222c.

(48) Wang, Y.-T.; Liu, W.; Li, Z.-P.; Yu, S.; Ke, Z.-J.; Meng, Y.; Tang, J.-S.; Li, C.-F.; Guo, G.-C. A Bubble-Induced Ultrastable and Robust Single-Photon Emitter in Hexagonal Boron Nitride. *arXiv:1906.00493* **2019**.

(49) Mendelson, N.; Morales, L.; Li, C.; Ritika, R.; Nguyen, M. A. P.; Loyola-Echeverria, J.; Kim, S.; Gotzinger, S.; Toth, M.; Aharonovich, I. Grain Dependent Growth of Bright Quantum Emitters in Hexagonal Boron Nitride. *arXiv:2005.10699* **2020**.

(50) Hayee, F.; Yu, L.; Zhang, J. L.; Ciccarino, C. J.; Nguyen, M.; Marshall, A. F.; Aharonovich, I.; Vučković, J.; Narang, P.; Heinz, T. F.; Dionne, J. A. Revealing Multiple Classes of Stable Quantum Emitters in Hexagonal Boron Nitride with Correlated Optical and Electron Microscopy. *Nat. Mater.* **2020**, *19*, 534–539. https://doi.org/10.1038/s41563-020-0616-9.

(51) Moon, H.; Bersin, E.; Chakraborty, C.; Lu, A.-Y.; Grosso, G.; Kong, J.; Englund, D. Strain-Correlated Localized Exciton Energy in Atomically Thin Semiconductors. *ACS Photonics* **2020**, *7* (5), 1135–1140. https://doi.org/10.1021/acsphotonics.0c00626.

(52) Zhou, X.; Zhang, Z.; Guo, W. Dislocations as Single Photon Sources in Two-Dimensional Semiconductors. *Nano Lett.* **2020**, *20* (6), 4136–4143. https://doi.org/10.1021/acs.nanolett.9b05305.

(53) Hohenberg, P.; Kohn, W. Inhomogeneous Electron Gas. *Phys. Rev.* **1964**, *136* (3B), B864–B871. https://doi.org/10.1103/physrev.136.b864.

(54) Kohn, W.; Sham, L. J. Self-Consistent Equations Including Exchange and Correlation Effects. *Phys. Rev.* **1965**, *140* (4A), A1133–A1138. https://doi.org/10.1103/physrev.140.a1133.

(55) Hamann, D. R. Optimized Norm-Conserving Vanderbilt Pseudopotentials. *Phys. Rev. B* **2013**, *88* (8), 085117. https://doi.org/10.1103/physrevb.88.085117.

(56) Schlipf, M.; Gygi, F. Optimization Algorithm for the Generation of ONCV Pseudopotentials. *Comput. Phys. Commun.* **2015**, *196*, 36–44. https://doi.org/10.1016/j.cpc.2015.05.011.

(57) Andreussi, O.; Brumme, T.; Bunau, O.; Nardelli, M. B.; Calandra, M.; Car, R.; Cavazzoni, C.; Ceresoli, D.; Cococcioni, M.; Colonna, N.; Carnimeo, I.; Corso, A. D.; Gironcoli, S. de; Delugas, P.; DiStasio, R.; Ferretti, A.; Floris, A.; Fratesi, G.; Fugallo, G.; Gebauer, R.; Gerstmann, U.; Giustino, F.; Gorni, T.; Jia, J.; Kawamura, M.; Ko, H.-Y.; Kokalj, A.; Küçükbenli, E.; Lazzeri, M.; Marsili, M.; Marzari, N.; Mauri, F.; Nguyen, N. L.; Nguyen, H.-V.; Otero-de-la-Roza, A.; Paulatto, L.; Poncé, S.; Giannozzi, P.; Rocca, D.; Sabatini, R.; Santra, B.; Schlipf, M.; Seitsonen, A. P.; Smogunov, A.; Timrov, I.; Thonhauser, T.; Umari, P.; Vast, N.; Wu, X.; Baroni, S. Advanced Capabilities for Materials Modelling with Quantum ESPRESSO. *J. Condens. Matter Phys.* **2017**, *29* (46), 465901. https://doi.org/10.1088/1361-648x/aa8f79.

(58) Perdew, J. P.; Burke, K.; Ernzerhof, M. Generalized Gradient Approximation Made Simple. *Phys. Rev. Lett.* **1996**, *77* (18), 3865–3868. https://doi.org/10.1103/physrevlett.77.3865.

(59) Liu, Y.; Zou, X.; Yakobson, B. I. Dislocations and Grain Boundaries in Two-Dimensional Boron Nitride. *ACS Nano* **2012**, *6* (8), 7053–7058. https://doi.org/10.1021/nn302099q.

(60) Comtet, J.; Glushkov, E.; Navikas, V.; Feng, J.; Babenko, V.; Hofmann, S.; Watanabe, K.; Taniguchi, T.; Radenovic, A. Wide-Field Spectral Super-Resolution Mapping of Optically Active Defects in Hexagonal Boron Nitride. *Nano Lett.* **2019**, *19* (4), 2516–2523. https://doi.org/10.1021/acs.nanolett.9b00178.

(61) Dietrich, A.; Bürk, M.; Steiger, E. S.; Antoniuk, L.; Tran, T. T.; Nguyen, M.; Aharonovich, I.; Jelezko, F.; Kubanek, A. Observation of Fourier Transform Limited Lines in Hexagonal Boron Nitride. *Phys Rev B* **2018**, *98* (8), 081414. https://doi.org/10.1103/physrevb.98.081414.

(62) Hossain, F. M.; Doherty, M. W.; Wilson, H. F.; Hollenberg, L. C. L. Ab Initio Electronic and Optical Properties of the N − V⁻ Center in Diamond. *Phys. Rev. Lett.* **2008**, *101* (22), 226403. https://doi.org/10.1103/physrevlett.101.226403.

(63) Dev, P. Fingerprinting Quantum Emitters in Hexagonal Boron Nitride Using Strain. *Phys. Rev. Res.* **2020**, *2* (2), 022050. https://doi.org/10.1103/physrevresearch.2.022050.

(64) Kim, K.; Lee, Z.; Malone, B.; Chan, K. T.; Alemán, B.; Regan, W.; Gannett, W.; Crommie, M. F.; Cohen, M. L.; Zettl, A. Multiply Folded Graphene. *Phys. Rev. B* **2010**, *83* (24), 245433. https://doi.org/10.1103/physrevb.83.245433.

(65) Wang, G.; Dai, Z.; Xiao, J.; Feng, S.; Weng, C.; Liu, L.; Xu, Z.; Huang, R.; Zhang, Z. Bending of Multilayer van Der Waals Materials. *Phys. Rev. Lett.* **2019**, *123* (11), 116101. https://doi.org/10.1103/physrevlett.123.116101.

**TABLE OF CONTENTS GRAPHIC**

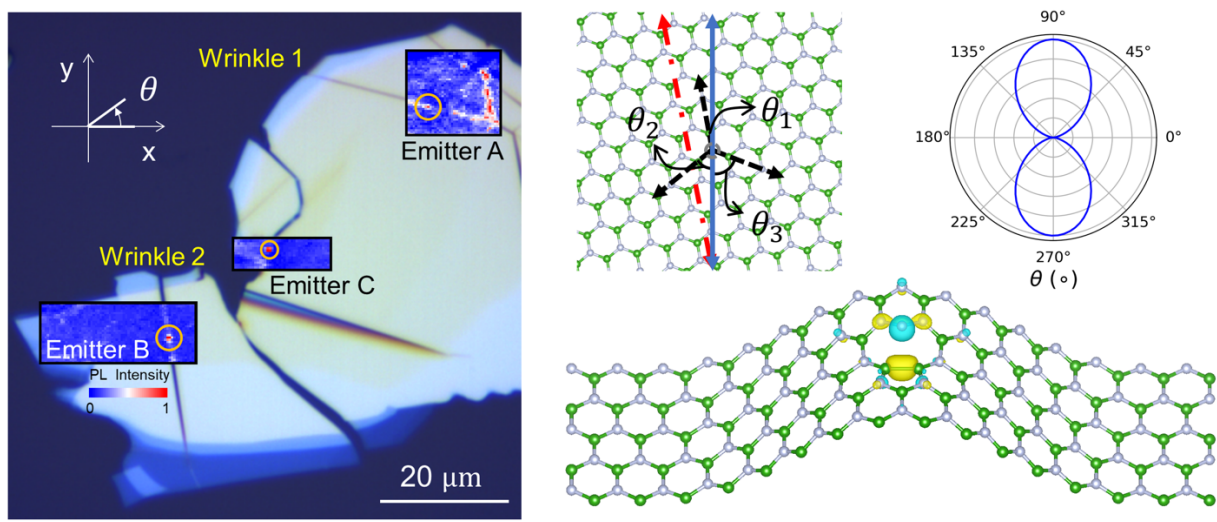

# Supporting Information of
# "Polarization and localization of single-photon emitters in hexagonal boron nitride wrinkles"

Donggyu Yim, Mihyang Yu, Gichang Noh, Jieun Lee*, and Hosung Seo*

*Department of Physics and Department of Energy Systems Research, Ajou University, Suwon, Gyeonggi 16499, Korea*

(*Correspondence to: jelee@ajou.ac.kr, hseo2017@ajou.ac.kr)

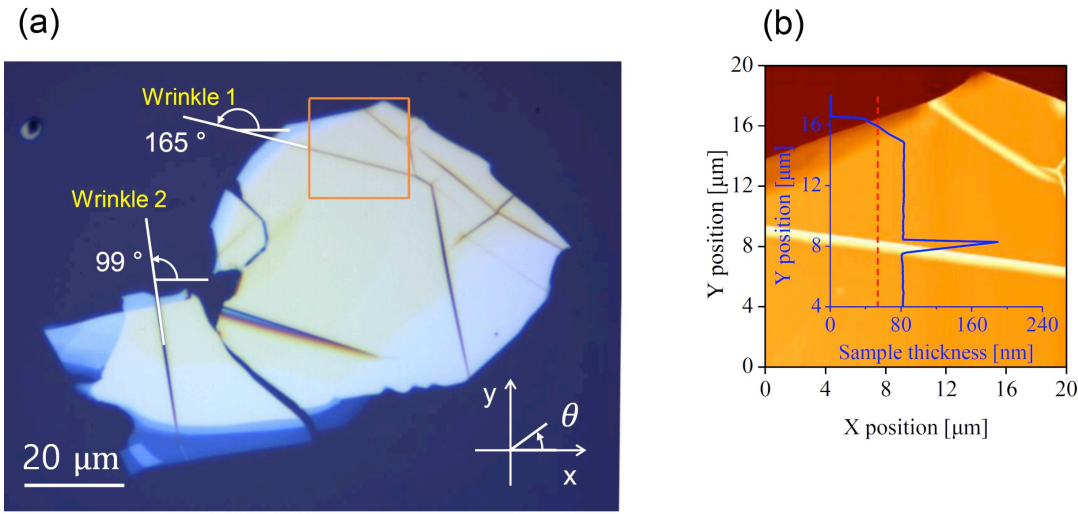

**Figure S1. Atomic force microscopy (AFM) of wrinkled h-BN.** (a) Microscope image of the h-BN flake used in our experiment. The angles of the wrinkle 1 and 2 measured from the *x*-axis are also shown. (b) AFM scan image on the area indicated by the orange box shown in (a). Inset shows the 1D measurement of the sample thickness measured along the red dashed line. The h-BN flake thickness is found to be 80 nm. The thickness and width of the wrinkle 1 is about 110 nm and 430 nm, respectively.

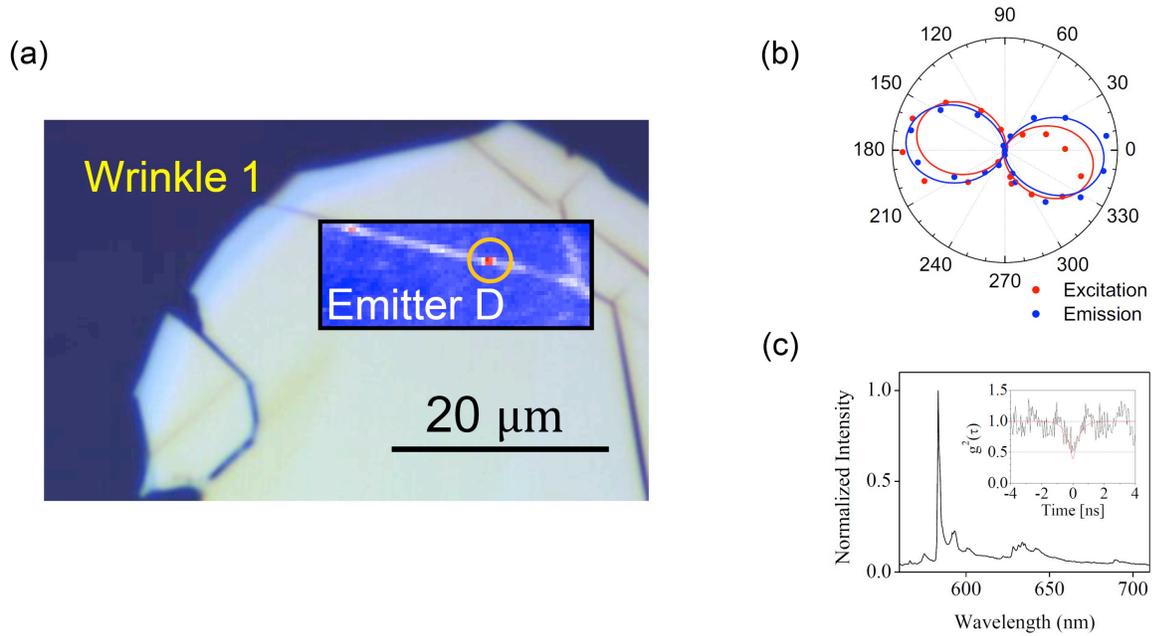

**Figure S2. Detailed measurement results of additional wrinkle-associated h-BN emitter.** (a) Microscope image of the h-BN flake and the 2D PL map of the emitter D overlapped on the image. (b) Polar plot of the excitation and emission polarization dependence of the emitter D on wrinkle 1. The polar plots are fitted by $I(\theta) = A\cos^2(\theta - \theta_0) + B$ to find the polarization angle, $\theta_0$. (c) The emission spectrum of the emitter D. In the inset, the second-order correlation measurement data (black solid line) is fitted by $g^2(\tau) = 1 - e^{-|\tau|/\tau_1}$ convoluted with the detector's response function (red solid line) where the emitter's lifetime $\tau_{1,D} = 0.39$ ns.

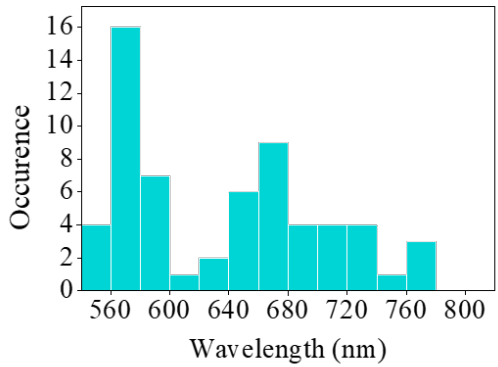 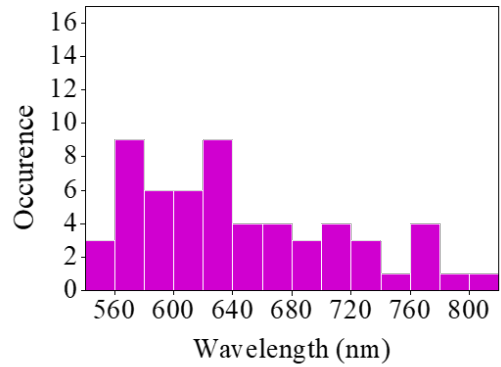

**Figure S3. The histogram of all observed emitters counted from 2D photoluminescence scan map.** (a) Emitters found on wrinkles. (b) Emitters found on the flat surface. Data from all observed emitters on the flake shown in Figure 1(a) are included. Some of the emitters disappeared after measurement. The statistics shows that emitters on the flat surface are found to have a wider spectral distribution.

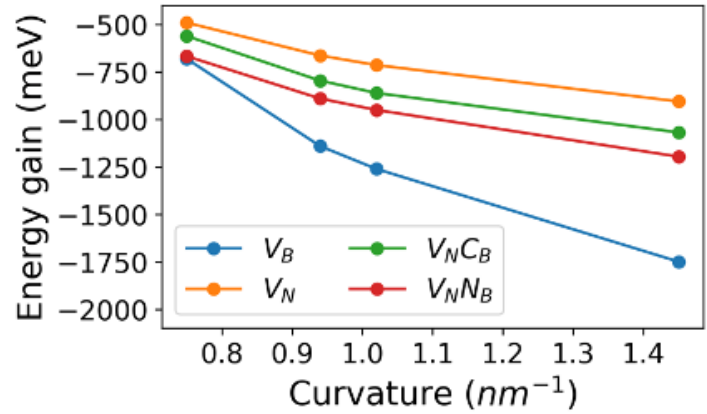

**Figure S4. The energy gain of vacancy-derived defects as function of curvature.** The curvature-driven energy gain of $V_N$, $V_B$, $V_NC_B$ and $V_NN_B$ formed on top of a wrinkle as a function of curvature.

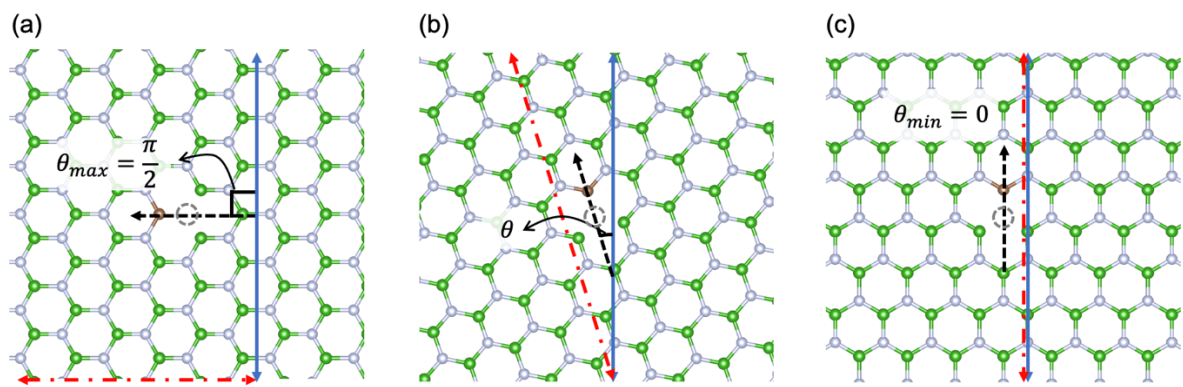

**Figure S5. Possible orientations of axial point defects in h-BN wrinkles.** (a) The maximum possible defect angle (90°) between a $V_NC_B$ defect and a wrinkle. $V_N$ and $C_B$ are denoted with a dotted circle and a brown ball, respectively. The defect axis is indicated by a black dashed arrow. The wrinkle direction is indicated by a solid arrow, which is perpendicular to the crystallographic arm-chair direction, which is denoted by a red dotdahsed arrow. (b,c) An intermediate possible defect angle (b) and the smallest possible defect angle (0°) (c) between a $V_NC_B$ defect and a wrinkle.

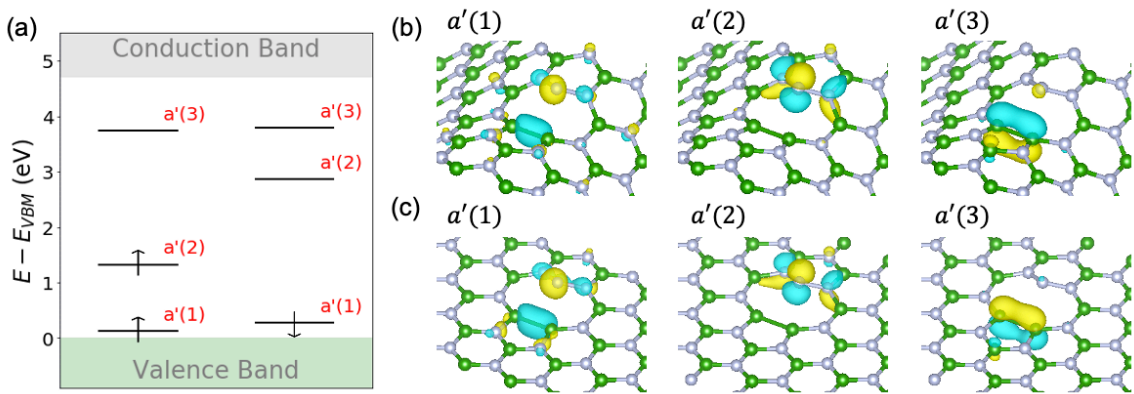

**Figure S6. Electronic structure and defect states of $V_NN_B$ defects** (a) Defect level diagram of $V_NN_B$ on the h-BN wrinkle. (b,c) Kohn-sham orbitals of the $V_NN_B$ defect on (b) and off (c) wrinkle. The occupied $a'(1)$ defect orbital is a $\sigma$-bonding state between boron dangling bonds, leading to dimerization of the B atoms.